\documentclass[11pt,twoside]{article}
\usepackage{asp2010}

\resetcounters

\begin{document}

\title{Halo Pairs in the Millennium Simulation: Love \& Deception}
\author{Jorge Moreno$^1$
\affil{$^1$SISSA, Astrophysics Sector, via Bonomea 265, 34136, Trieste, Italy}
}

\begin{abstract}
In this work I investigate the statistical properties of a huge catalog of closely interacting pairs of dark matter haloes, extracted from the Millennium Simulation database.  Only haloes that reach a minimum mass $\geq 8.6 \times 10^{10} M_{\odot}\, h^{-1}$ (corresponding to 100 particles) are considered.  Close pairs are selected if they come within a critical distance $d_{\rm crit}$.  I explore the effects of replacing $d_{\rm crit}=1\, {\rm Mpc}\, h^{-1} \rightarrow 200\, {\rm kpc}\, h^{-1}$ on the evolution of separations, lifetimes, total masses and mass ratios of these pairs.
\end{abstract}

\section{Introduction}

Mergers of galaxies play a fundamental role in essentially all modern theories of galaxy formation.  They are believed to determine the morphology of a galaxy, drive its star formation and even activate its nuclear supermassive black hole.  This has motivated the development of very detailed numerical simulations of merging galaxies (often involving only two galaxies in isolation).  Unfortunately, with very few exceptions \cite[e.g.,][]{tonnesen11}, these simulations typically concentrate primarily on the post-merger aftermath, often neglecting the early stages of interaction.  

Observations, on the other hand, tell us a different story.  Surveys like the SDSS and zCOSMOS have confirmed, in large numbers, that galaxies in pairs tend to be bluer, have their star formation enhanced, and are more likely to be active \citep[e.g.,][]{ellison11,silverman11}.  Moreover, the discovery rate of binary quasars has accelerated to unprecedented levels in the last few years \citep[e.g.,][]{liu11}.  For this reason, it is vital to re-focus our attention to the early stages of galactic interactions.

\section{The Halo-Pair Catalog}

This work is based on the results of Moreno (2011), where I use the publicly available Millennium Simulation database \citep{springel05}  to construct a very large catalog of closely-interacting pairs of haloes.  First, pairs are selected if they come within a critical distance $d_{\rm crit}$.  Also, I require that all haloes reach a mass of at least $\geq 8.6 \times 10^{10} M_{\odot}\, h^{-1}$ (corresponding to 100 particles). Below I explore the effects of replacing $d_{\rm crit}=1\, {\rm Mpc}\, h^{-1} \rightarrow 200\, {\rm kpc}\, h^{-1}$ (these two sets are named the `Close Set' and the `Very Close Set' respectively).  The true physical separations at which interaction-induced phenomena are instigated may depend on the environment and on the actual phenomenon we care about. For this reason, at this preliminary stage, I choose the simplest possible criterion.  More sophisticated conditions can always be implemented, and the corresponding catalogs can then be extracted from this primary set.

To describe the evolution of each pair, I propose a scenario based on {\bf five-stages}:  {\it initial}, {\it entry}, {\it closest}, {\it final} and {\it fate}.  The initial, closest and final stages are those at which the two haloes are identified in the simulation for the first time, at their closest distance, and for the last time respectively.  The entry stage is the first time the pair is at a distance lower than $d_{\rm crit}$.  The fate stage comes in two flavors:  either the two haloes merge or not.  The latter situation happens because either they do not have enough time to merge by the present time or one of the haloes is absorbed by a third halo.  This process is analogous to a situation that occurs commonly in real life.  Suppose you meet a person and the two of you date for a while.   The following two things can happen:  either you marry this person eventually ({\it love}) or someone else takes this person away from you ({\it deception}).  In the simulation, the culprit of this deception is usually a very massive halo in the vicinity.  Figure~1 illustrates these two situations.

\begin{figure}
    \centering
    \includegraphics[width=55 mm]{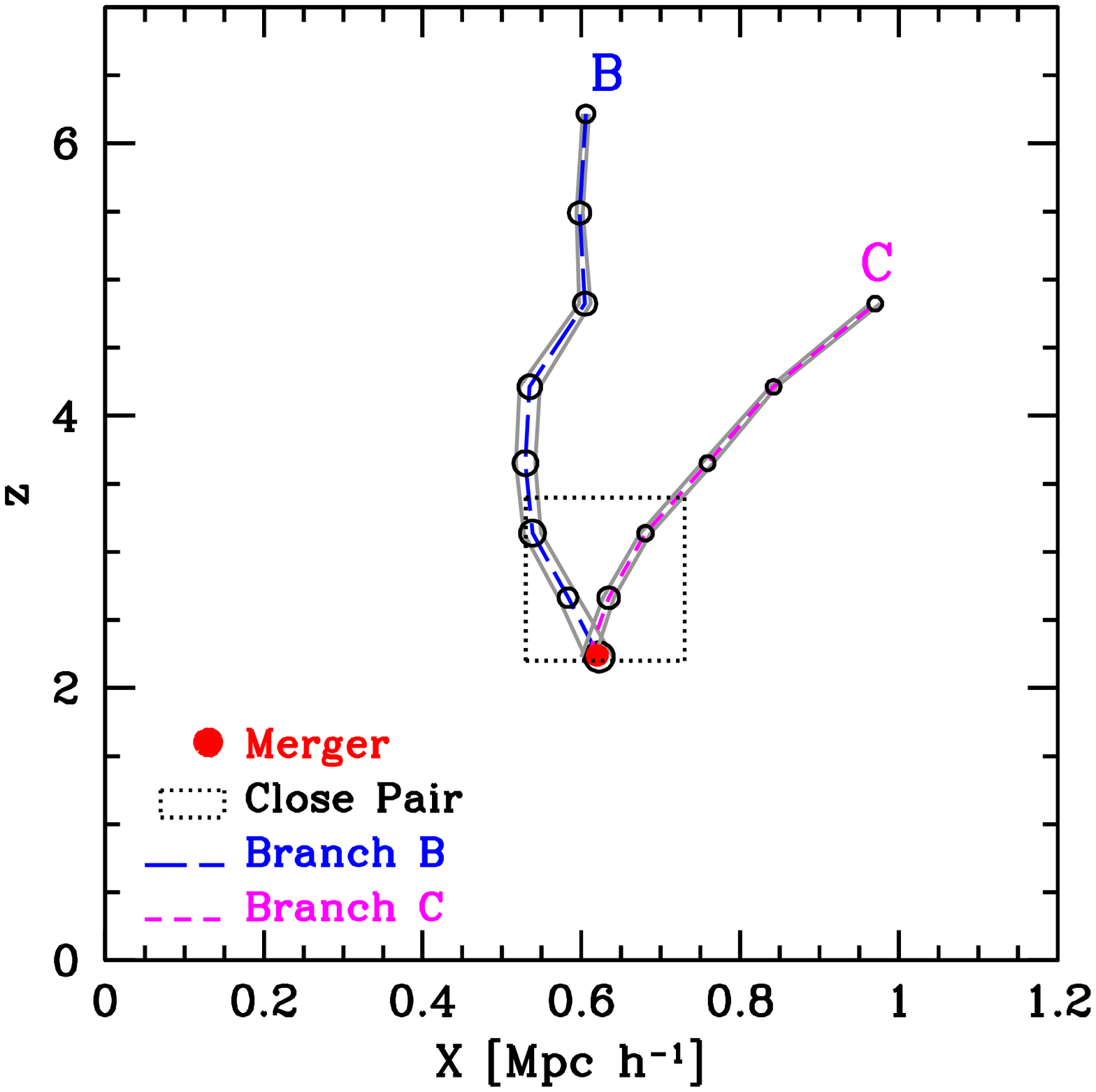}
     \includegraphics[width=55 mm]{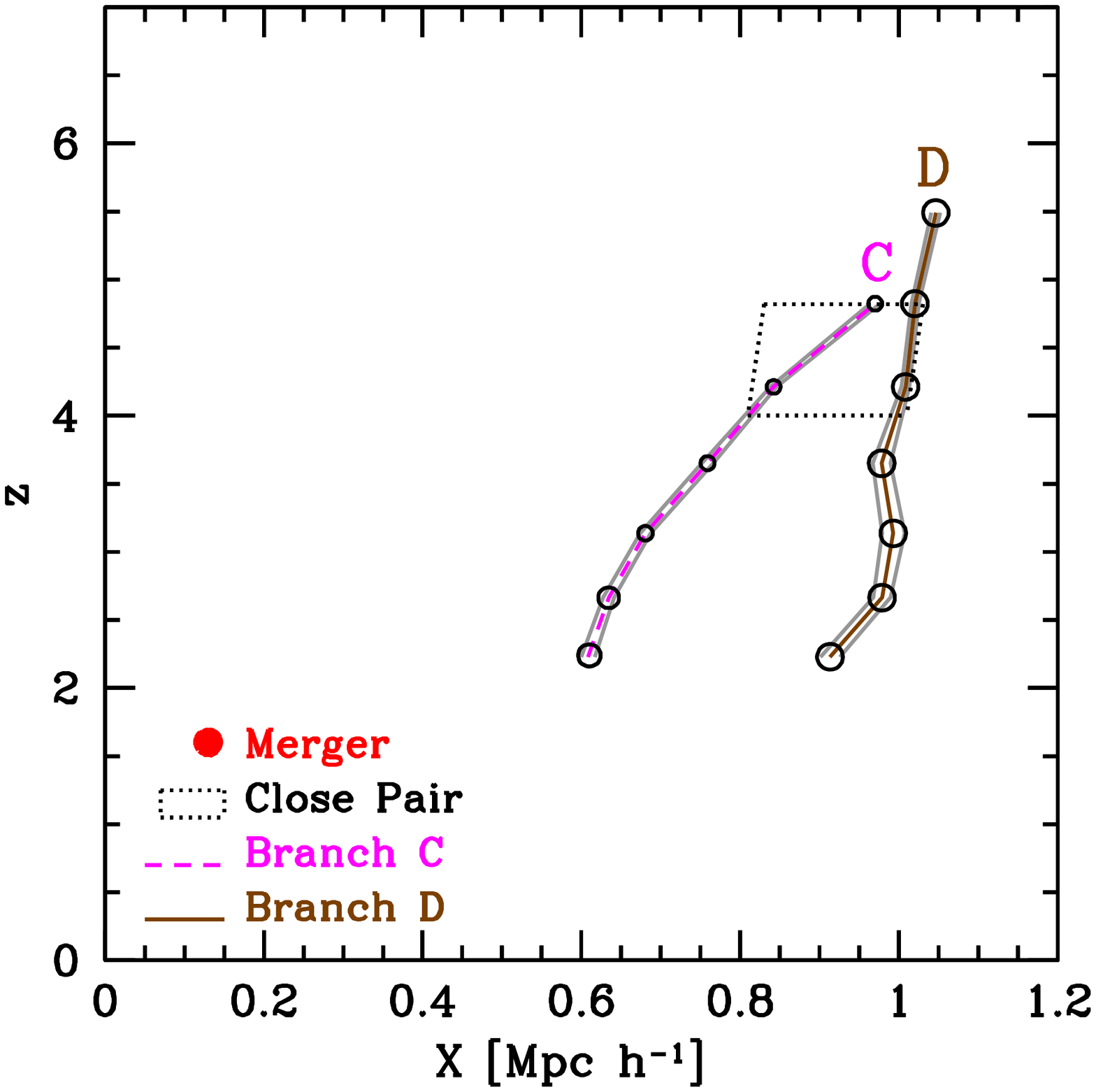}
\caption{Pairs of branches extracted from the merger tree presented in Figure 1 of \cite{moreno11}.  Red solid circles represent mergers, while dotted parallelograms refer to close interactions.  LEFT-PANEL:  Branches B and C merge ({\it love}).  RIGHT-PANEL:  Branches C and D interact for a while, but split eventually ({\it deception}).} 
    \label{fig1}
\end{figure}

\section{Results}

For the sake of readability, I present all my results together in Figure~2.  The corresponding colors and line-styles are indicated in the key.  To show the effect of  shrinking $d_{\rm crit}=1\,{\rm Mpc}\,h^{-1} \rightarrow 200\,{\rm kpc}\,h^{-1}$, I split each of the four panels in two:  the upper portions refer to the Close Set, while the lower portions correspond to the Very Close Set.  Before going into details, the first thing to notice is that when a smaller $d_{\rm crit}$ is reduced, halo pairs that do {\it not} merge are less likely to be selected (from $53\% \rightarrow 42\%$).

Halo pairs usually begin at very high redshift with {\bf separations} $\sim10$ Mpc $h^{-1}$.   They cross the critical threshold, reach their minimum separation, and are identified as two distinct objects for the final time.  While a fraction of these pairs end up with sub-critical separations, many of these have super-critical final separations.  This fraction increases as $d_{\rm crit}$ is reduced (from $45\% \rightarrow 57\%$).  In other words, using a smaller critical distance selects pairs that end up with more elongated orbits.  For those haloes that never merge, shrinking $d_{\rm crit}$ actually augments the fraction of super-critical fate separations (from $40\% \rightarrow 80\%$).  In other words, while some super-critical pairs at the final stage come back and merge, others separate permanently.  Using a smaller critical distance enhances the contribution of cases where this splitting is more violent.

\begin{figure}
    \centering
    \includegraphics[width=48 mm]{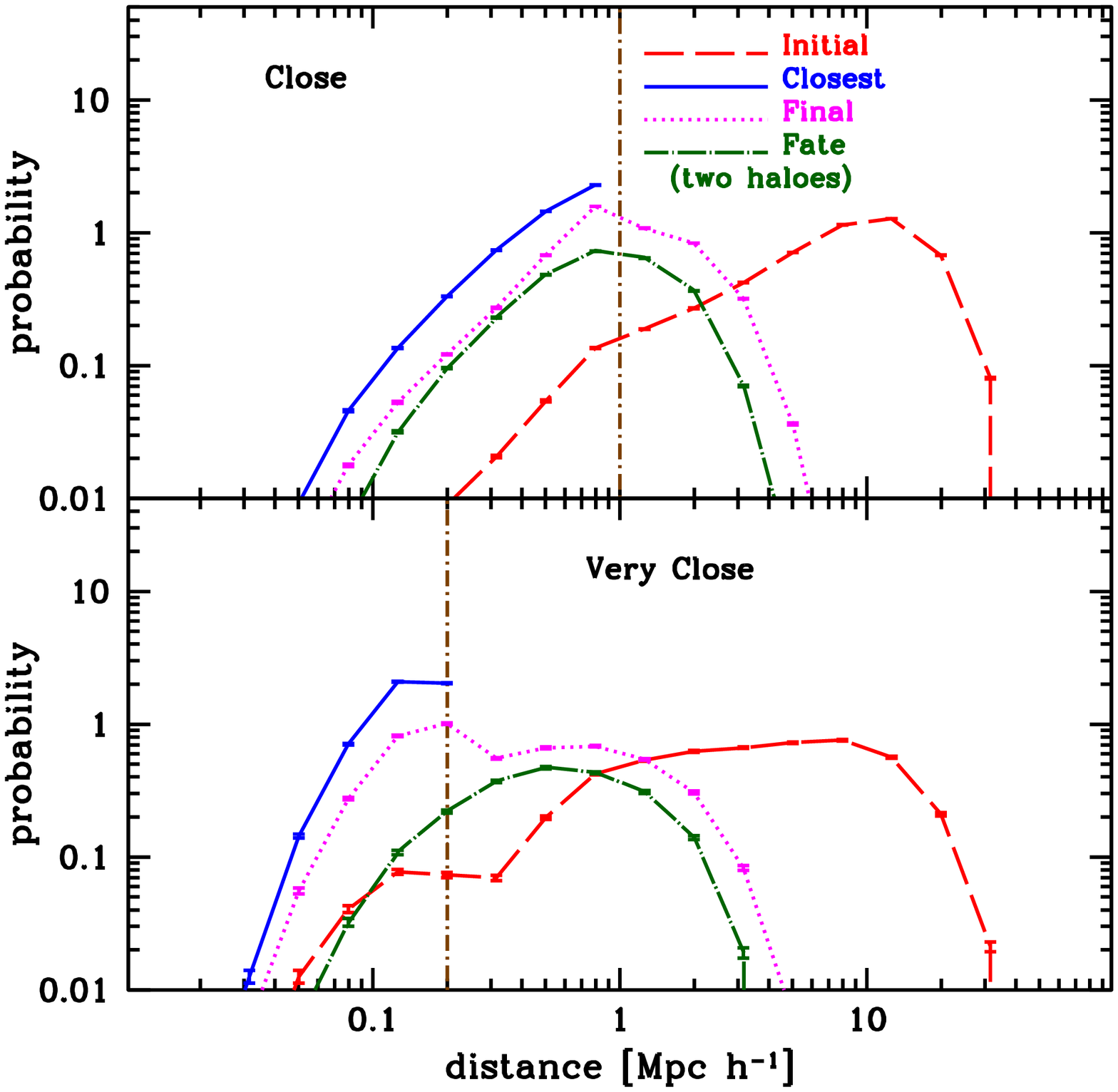}
     \includegraphics[width=48 mm]{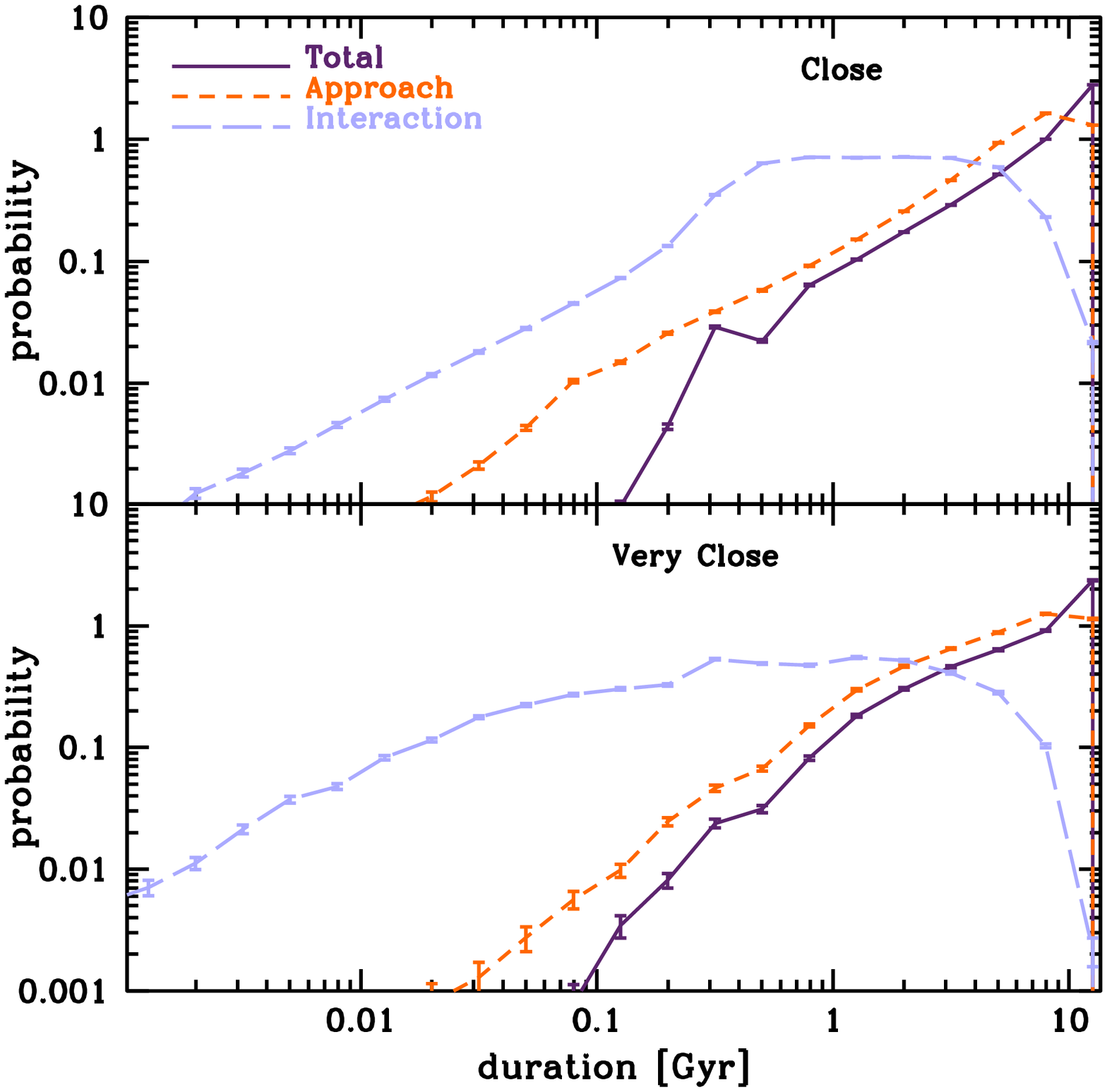}
       \includegraphics[width=48 mm]{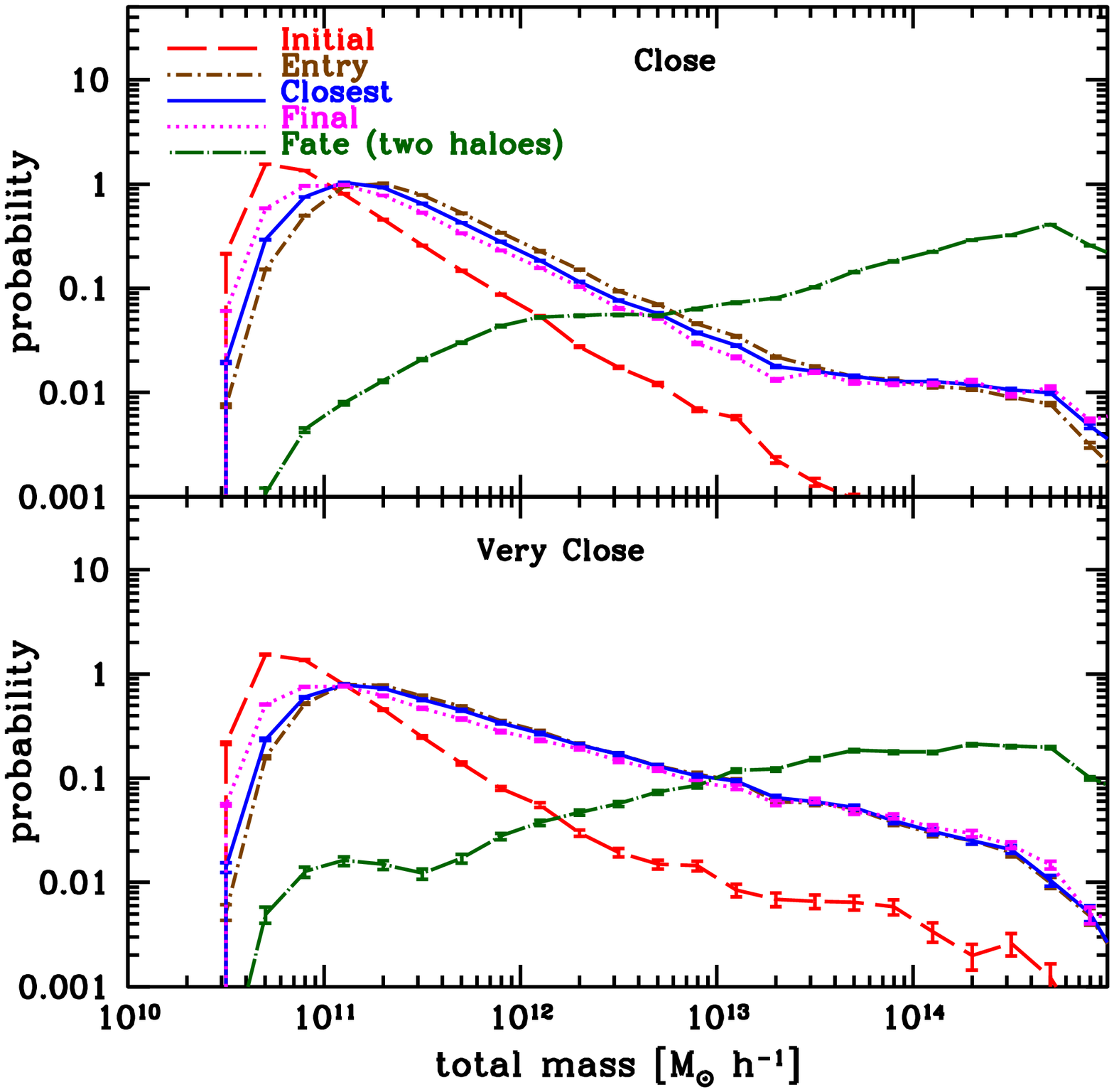}
     \includegraphics[width=48 mm]{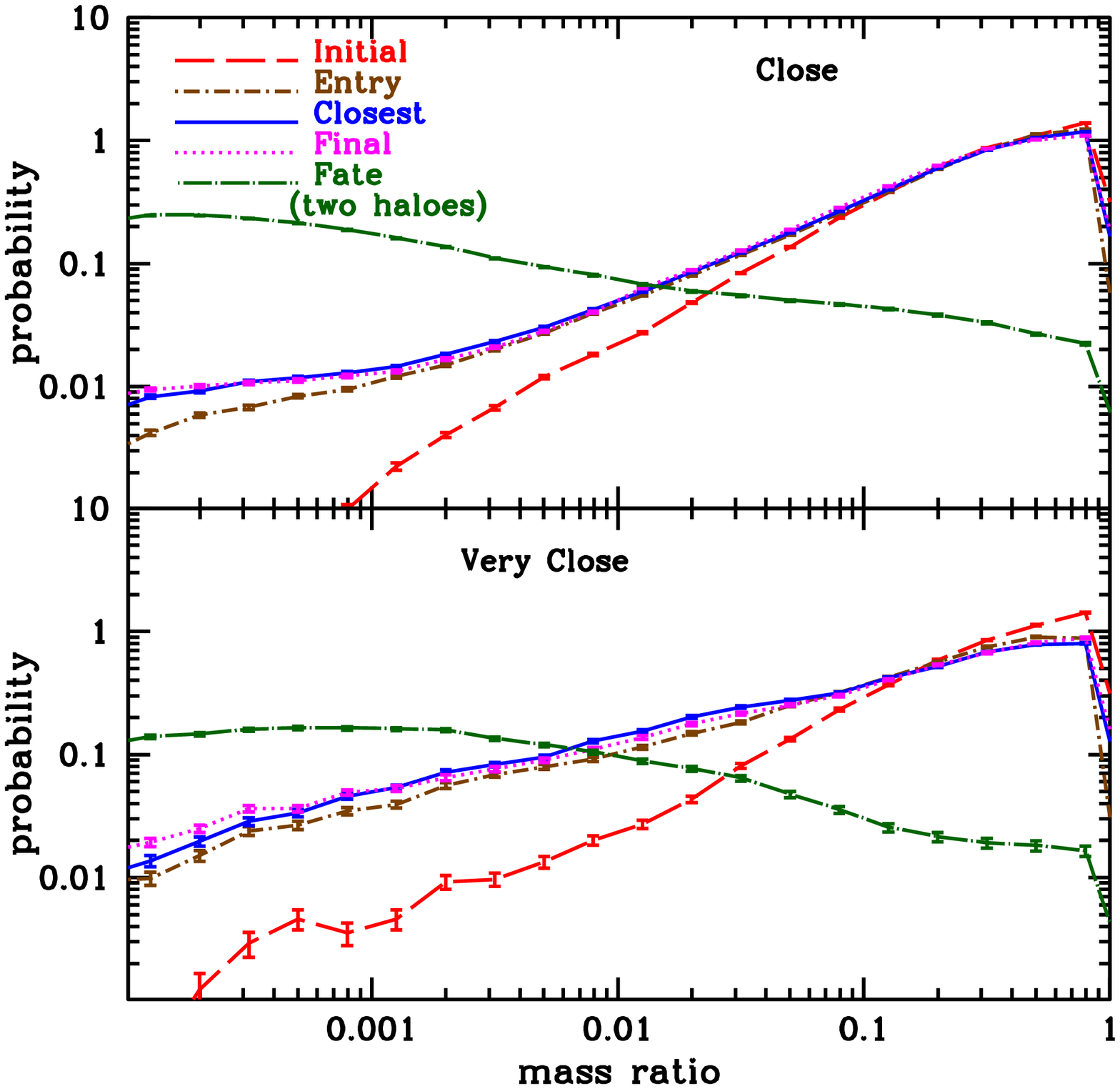}
\caption{Results.  Each of the four panels is split in two:  the upper portions refer to $d_{\rm crit}=1\,{\rm Mpc}\,h^{-1}$ (Close Set), while the lower portions refer to $d_{\rm crit}=200\,{\rm kpc}\,h^{-1}$ (Very Close Set).  UPPER-LEFT:  Separation.  UPPER-RIGHT:  Lifetime.  LOWER-LEFT:  Total mass $m_1+m_2$.  LOWER-RIGHT:  Mass ratio $\mu=m_1/m_2$ (with $m_1\leq m_2$).} 
    \label{fig2}    
\end{figure}

The {\it total} {\bf lifetime} of the pair is the sum of the {\it approaching time} (from initial to entry) and the {\it interacting time} (from the first to the last time the separation is below $d_{\rm crit}$).  The total duration of the pair can be very long, almost always larger than a few Gigayears.  In fact, most of this time is spent in the approaching phase.  The distribution of the total and infall times is largely insensitive to the choice of critical distance.  However, the interaction times tend to be shorter when $d_{\rm crit}$ is reduced (e.g., the fraction of pairs with interactions shorter than 1 Gyr goes from $40\% \rightarrow 60\%$).

Haloes in pairs typically begin with {\bf masses} just above the resolution of the simulation.  As time progresses, the sum of the masses tends to increase at scales $\geq 3 \times 10^{13} M_{\odot} h^{-1}$, otherwise they decrease.  In other words, this particular mass scale marks the transition at which the evolution is dominated by tidal shredding as the pair sinks within a larger host.  The great majority of pairs remain in the small-mass regime throughout their existence.  Reducing $d_{\rm crit}$ increases the fraction of pairs in the massive regime (from $3\% \rightarrow 8\%$ during the interaction phase).  In contrast, the fate mass distribution is very different from the others because it is largely determined by external haloes in the vicinity.  Notice that the fate total masses tend to be extremely large!

Since the initial masses are usually near the resolution of the simulation, pairs tend to be born with similar masses (and thus with {\bf mass ratios} $\mu \lesssim 1$; e.g., in the `major' pair regime).  Throughout their evolution, pairs tend to stay more or less major; and only at the interaction stages do the more discrepant `minor' pairs begin to appear.  
The importance of minor pairs becomes more prevalent as $d_{\rm crit}$ decreases (from $19\% \rightarrow 33\%$ for $\mu \leq 1/10$). Nevertheless, minor pairs never dominate the bulk of the population.  In contrast, notice that fate mass ratios tend to be extremely small (i.e., very discrepant)! 

In particular, the total-mass and mass-ratio distributions point to the fact that at small scales (probed with the smaller $d_{\rm crit}$), the fraction of pairs involving a very massive member increases.  Moreover, these results in conjunction also appear to unveil the real culprits responsible for splitting close pairs: very massive haloes in the vicinity (usually the central halo in a group or some other massive secondary halo) absorb one of the members of the pair before these two have a chance to merge.  And as smaller scales are probed, the effects of these vicious culprits become even more evident.

\section{Conclusions}

Using the Millennium Simulation, I analyze the behavior of a very large set of halo pairs selected by a proximity criterion.  As this critical distance is reduced, (1) more mergers are pre-selected, (2) pairs end up with more elongated orbits, (3) interactions tend to be shorter, and (4) non-mergers split more violently.  Using a smaller $d_{\rm crit}$ probes smaller scales, enhancing the importance of pairs involving a massive member and the prevalence of massive neighbors capable of splitting pairs.

With these results, I wish to highlight the following important lessons:
\begin{itemize}
\item {\bf Extra care must be taken when using close galaxy pairs as proxies for mergers. This subtlety is important even in physical three-dimensional space!}

\item {\bf Systems with two merging galaxies in isolation are just an approximation.  In reality, the universe can nurture these duos, or split them altogether!}

\end{itemize}

Next I will use this catalog to explore the symbiosis between interacting galaxies and binary quasars.  Ultimately, this will set up realistic initial conditions for black hole mergers.  This work is just a puzzle piece of an ambitious long-term research program centered on the evolution of galaxies, supermassive black holes, and their environment.

\acknowledgements   I thank the organizers of the {\it Galaxy Mergers in an Evolving Universe} for such a lovely conference.  To everyone in the field who believes in me and keeps encouraging me -- you know who you are!  Lastly, to all my fellow participants for helping me perpetuate my theory of marriage -- and extra thanks to Brad Pitt for inspiring this research. This work is funded by SISSA Young Scientist Grant ASTR 639 and CONACyT -- and is dedicated to my wife Viridiana: just love, no deception!

\bibliography{moreno}

\end{document}